# Automatic Mapping Tasks to Cores - Evaluating AMTHA Algorithm in Multicore Architectures

Laura De Giusti[1], Franco Chichizola[1], Marcelo Naiouf[1], Armando De Giusti[1], Emilio Luque[2]

[1] Instituto de Investigación en Informática (III-LIDI) – School of Computer Sciences – Universidad Nacional de La Plata
La Plata, Buenos Aires, Argentina

[2] Computer Architecture and Operating System Departtment (CAOS) - Universidad Autónoma de Barcelona
Barcelona, España

**Abstract**
The AMTHA (Automatic Mapping Task on Heterogeneous Architectures) algorithm for task-to-processors assignment and the MPAHA (Model of Parallel Algorithms on Heterogeneous Architectures) model are presented.
The use of AMTHA is analyzed for multicore processor-based architectures, considering the communication model among processes in use.
The results obtained in the tests carried out are presented, comparing the real execution times on multicores of a set of synthetic applications with the predictions obtained with AMTHA.
Finally current lines of research are presented, focusing on clusters of multicores and hybrid programming paradigms.

## 1. Introduction

A cluster is a parallel processing system formed by a set of PCs interconnected over some kind of network and that cooperate as if they were an "only and integrated" resource, regardless of the physical distribution of its components. When two or more clusters are connected over a LAN or WAN network, we are in the presence of a multicluster [1].

The hardware and operating system of the participating machines may be different; each machine may even be a "multiprocessor", as is the case in multicore architectures that are so relevant nowadays. Multicore processors include several processing elements within an integrated circuit. This type of architectures are considered as a solution to the limitation of one core machines to increase computing power due to the increase in temperature [1][2][3].

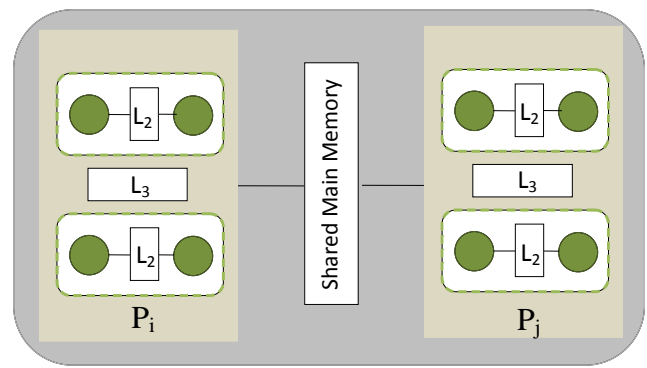

Fig. 1. Diagram of a multicore structure.

Figure 1 shows the typical design of a current multicore architecture, composed by two processors that share the main memory. Each of these processors in turn is formed by four cores that share one $L_3$ cache memory (this memory may not be present in some models). There is also an $L_2$ cache memory that is shared by pairs of cores. Finally, each core has its own $L_1$ cache memory.

As it can be seen from Fig. 1, the communication between the various cores is done through the different memories of the architecture. Thus, the communication time between two cores is given by the time required to access the corresponding memory. In the case of Fig. 1, there are three levels of shared memory with their corresponding communication times.

It is possible to build clusters (also multiclusters) using multicores; Fig. 2 shows a diagram of this type of architecture, where each $P_i$ may have $K$ processing cores and $L$ memory levels (in the example of Fig. 1, $K=4$ and $L=3$).

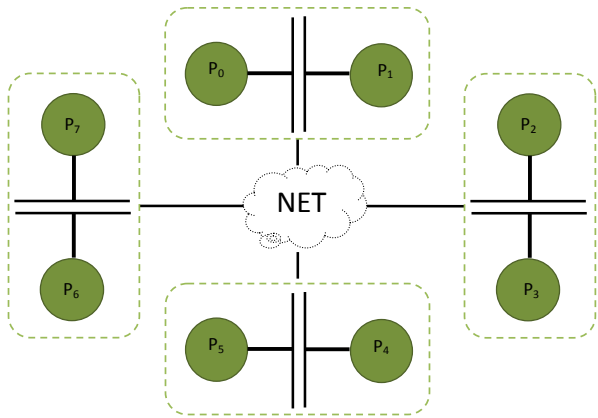

Fig. 2. Diagram of a multicore cluster.

The advent of this type of distributed architectures that can be accessed by any user has promoted the growth of parallel processing as a technique to increase architecture exploitation. Application programmers should implement this technique by describing its component tasks and the interaction between them. Once the application has been developed, the programmer or user of the application on a parallel architecture will have to decide how to do it. That is, they should select how many of the processors/cores in the architecture will be used and how the application tasks will be assigned to them, in order to achieve the best possible throughput of the architecture with the lowest response time. This problem of solving the distribution of tasks between processors is called scheduling. [4]

The problem of the automated scheduling of tasks to processing elements (processors in conventional machines and cores in multicore computers) is highly complex [5]. This complexity can be briefly represented considering the two main elements relating the parallel application to the supporting architecture: each node's processing capacity and the cost in time of communicating two processing elements [6].

The goal of modeling processing architectures is to obtain an "abstract" or simplified version of the physical machine, capturing crucial characteristics and disregarding minor details of the implementation [7].

A model does not necessarily represent a given real computer, but allows studying classes of problems over classes of architectures represented by their essential components. In this way, a real application can be studied over the architecture model, allowing us to get a significant description of the algorithm, draw a detailed analysis of its execution, and even predict the performance [8].

In the case of parallel systems, the most currently used architectures – due to their cost/performance relation - are clusters and multiclusters of multicores; for this reason, it is really important to develop models fitting the characteristics of these platforms. Essential elements to be considered are the potential heterogeneity of processors/cores, the communication resources (shared memory, buses, networks), and the different cache levels, which add complexity to the modeling [9] [10].

At present, there are different graph-based models to characterize the behavior of parallel applications in distributed architectures [11]. Among these models, we can mention TIG (Task Interaction Graph), TPG (Task Precedence Graph), TTIG (Task Temporal Interaction Graph) [12], TTIGHA (Task Temporal Interaction Graph on Heterogeneous Architectures) [13] and MPAHA (Model on Parallel Algorithms on Heterogeneous Architectures) [14].

Once the graph modeling the application has been defined, the *scheduling* problem is solved by an algorithm that establishes an automatic mechanism to carry out the assignment of tasks to processing elements, searching for the optimization of some running parameter (usually, time) [15][16][17]. Among the known mapping/scheduling algorithms, we consider AMTHA (Automatic Mapping Task on Heterogeneous Architectures), a mapping algorithm to carry out the assignment of tasks, making up the application to the processors of the architecture [14]. This algorithm considers the heterogeneous characteristics of the architecture taken into account in MPAHA (Model on Parallel Algorithms on Heterogeneous Architectures) model [14].

The AMTHA algorithm was developed to carry out the scheduling of applications executed over cluster and multicluster architectures using conventional machines with good comparative results [14].

## 2. Contribution

This paper focuses in the analysis of MPAHA model and AMTHA algorithm on multicore cluster architectures. The operation of the AMTHA mapping algorithm is tested over two different architectures with 8 and 64 cores.

In Section 3, the scheduling algorithm AMTHA and the MPAHA model are described. Section 4 deals with the possible use of AMTHA and MPAHA for multicore clusters. In Section 5, the experimental work carried out with a multicore machine is presented, and the results obtained are detailed in Section 6. Finally, Section 7 presents the conclusions and the future lines of work.

## 3. AMTHA mapping algorithm

AMTHA is a static mapping algorithm that is applied to the graph generated by the MPAHA model. It allows determining the assignment of tasks to the processors of the architecture to be used, minimizing the execution time of the application. This algorithm must also provide the order in which subtasks (forming the task) assigned to each processor should be executed (task scheduling) [14].

The MPAHA model is based on the construction of a directed graph G (V,E), where:
- *V* is the set of nodes representing each of the tasks $T_i$ of the parallel program. Each node represents a task $T_i$ of the parallel program, including its subtasks ($St_j$) and the order in which they should be executed to perform the task. If there is a heterogeneous architecture, the computation times for each processor should be taken into account ($V_i$ (s,p) = execution time of subtask *s* in processor type *p*).
- *E* is the set of edges representing each of the communications between the nodes of the graph. An edge *A* between two tasks $T_i$ and $T_j$ contains the communication volume (in bytes), the source subtask ($\in T_i$) and a target subtask ($\in T_j$).

It should be noted that, given the heterogeneity of the interconnecting network, instead of representing the time required for the communication, the corresponding communication volume between two subtasks is represented.

AMTHA considers an architecture with a limited number of heterogeneous processors. As regards the interconnecting network, the algorithm also considers that bandwidth and transmission speed can be heterogeneous.

The AMTHA algorithm uses the values of graph *G* generated by the MPAHA model; these values are the time required to compute a subtask in each type of processor, the communication volume with adjacent processors, and the task to which each subtask belongs.

The AMTHA algorithm assigns one task at a time until all tasks have been assigned. Figure 3 shows the pseudo-code with the main steps of the algorithm.

When the execution of the AMTHA algorithm ends, all the tasks have been assigned to one of the processors and the order in which the subtasks forming the tasks assigned to these processors will be executed has also been determined.

---
Calculate *rank* for each task.
Whereas (not all tasks have been assigned)
1. Select the next task *t* to assign.
2. Chose the processor *p* to which task *t* should be assigned.
3. Assign task *t* (selected in step 1) to processor *p* (selected in step 2).
4. Update the *rank* of the tasks involved in step 3.

---

Fig. 3. Pseudo-code with the basic steps of the AMTHA algorithm.

The following paragraphs describe each of the three steps followed during the execution of the AMTHA algorithm.

### 3.1. Calculating the rank of a task

Given a graph *G*, the rank of a task $Rk(T)$ is defined as the sum of the average times of the subtasks forming it and that are ready for execution (all predecessors have already been assigned to a processor and are already there). Eq. (1) expresses this definition:

$$Rk(T) = \sum_{i \in L(T)} W_{avg}(St_i) \quad (1)$$

$L(T)$ is the set of subtasks that are ready for task *T*. $W_{avg}(St)$ is the average time of subtask *St*. The average time is calculated as shown in Eq. (2).

$$W_{avg}(St_i) = \frac{\sum_{p \in P} V_{St_i}(type\ of\ processor\ p)}{\#P} \quad (2)$$

*P* is the set of processors present in the architecture and *#P* is the number of processors forming this set.

### 3.2. Selecting the task to execute

After obtaining the *rank* of each application task, the task that maximizes it is selected. If there are two or more tasks that have the same maximum value, the algorithm breaks this tie by selecting the one that minimizes the total execution time average for the task. Eq. (3) shows this calculation:

$$Tavg(T) = \sum_{i \in T} W_{avg}(St_i) \quad (3)$$

### 3.3. Selecting the processor

Selecting the processor involves choosing the computer within the architecture that minimizes the execution time when the selected task is assigned to that processor.

In order to understand how the time corresponding to processor *p* is calculated, the fact that each processor keeps a list of subtasks $LU_p$ that were already assigned to it and that can be executed (all its predecessors are already placed), and another list that contains those subtasks that were assigned to *p* but whose execution is still pending $LNU_p$ (some of their predecessors have not been placed yet) must be taken into account.

Therefore, to calculate which processor *p* will be selected, two possible situations are considered:
1. All subtasks of task *t* can be placed in *p* (that is, all its predecessors have been placed).
2. Some of the subtasks of *t* cannot be placed in *p* (this happens when some predecessor of this subtask has not been placed).

In the first case, the time *Tp* corresponding to processor *p* is given by the moment in which *p* finishes the execution of the last subtask of *t*. However, in the second case, the time *Tp* corresponding to processor *p* is given by the time when the last subtask of $LU_p$ will finish plus the addition of all execution times in *p* for each of the subtasks on $LNU_p$. This takes into account the synchronization and idle waiting times.

### 3.4. Assigning the task to the selected processor

When assigning a task *t* to a processor/core *p*, there is an attempt to place each subtask $St_k$ belonging to *t* to the processor at a moment in time when all the adjacent subtasks have already finished (including the predecessor subtask within *t*, if there is one) and its communications have been completed. The assignment can be a free interval between two subtasks that have already been placed in *p*, or an interval after them. If subtask $St_k$ cannot be placed, it is added to the $LNU_p$ list. Each time a subtask $St_k$ is added to the *LU* list of one of the processors, an attempt is made to place all the predecessors belonging to already assigned tasks.

### 3.5. Updating the rank value of the pending tasks.

The first action within this step consists in assigning *-1* as *rank* value to the task *t* that was assigned to processor *p*. The reason for this is to prevent task *t* from being re-selected for assignment.

Also, the following situation is considered in this step: for each subtask $St_k$ placed in step 3.4, the need to update the *rank* of the tasks to which successor subtasks $St_{succ}$ of $St_k$ belong is analyzed; that is, if all predecessors of $St_{succ}$ are already placed, then the *rank* of the task to which $St_{succ}$ belongs is updated by increasing it by $W_{avg}(St_{succ})$.

## 4. MPAHA and AMTHA in multicore clusters

### 4.1. MPAHA model

The MPAHA model described in Section 2 does not require any modification to be used with multicore processors or multicore clusters. The directed graph G (V,E) representing the tasks $T_i$ and the communication among them do not change, if the parallel program is the same; regardless of the physical architecture. This is coherent with the previous definition of "*model*".

### 4.2. AMTHA algorithm

When the AMTHA algorithm is run over a multicore cluster architecture, the following issues should be considered:

- The tasks that are part of the applications to execute will now be placed in some of the cores in the architecture; these cores may belong to any of the physical processors $P_i$ of Fig. 2.

- The heterogeneity of the architecture as regards communications is not only given by the existence of different interconnecting networks within the architecture, but also by the different memory levels (main or cache) shared by the cores within each multicore machine. That is, two cores of the global architecture may communicate through different levels of shared memory, or by means of messages sent through an interconnecting network, as can be seen in Fig 1.

- When the algorithm assigns a task, it must consider the communication costs with its predecessor tasks. To this aim, data related to the communication types that occur through the interconnecting network used when working with conventional clusters are required, as well as additional information regarding average access times for each of the memory levels in the multicores, together with information about core distribution in the machine.

## 5. Experimental work

In order to analyze the applicability of the AMTHA algorithm over multicore architectures, a set of synthetic applications with various characteristics was generated (as indicated in Section 5.1). For each of these, task assignment to the different cores in the architecture using the AMTHA algorithm was determined, and the execution time of using such distribution was estimated ($T_{est}$).

Based on the distribution done with AMTHA, the application was executed over the architecture described in Section 5.2 in order to obtain the real execution time ($T_{exec}$).

Both times ($T_{est}$ and $T_{exec}$) were compared to determine how well the AMTHA algorithm estimates the execution time.

### 5.1. Choosing the set of applications to evaluate the AMTHA algorithm

A set of applications was selected, in which each of them varied in terms of typical parameters: task size (5-50 seconds), number of subtasks making up a task (3-6), communication volume among subtasks (1000-10000), and communication probability between two different subtasks (5-35 %).

Initially we worked with 15-25 tasks (with 8 cores) and now we increased the number of tasks to 120-200, using 64 cores.

In all the applications, the total computing time exceeds that of communications (coarse grained application).

### 5.2. Choosing the architecture for the tests

Initial architecture was a Dell Poweredge 1950 with 2 quad core, 2.33 GHz Intel Xeon e5410 processors; 4 Gb of RAM (shared between both processors); 6 MB L2 cache for each pair of processor cores.

Actually we are working with HP BL260c G5 with 64 cores in 8 blades with 2 INTEL E5405 processors with a quad core configuration and 2 Gb of local RAM.

## 6. Results

To analyze the results of the tests carried out, the difference between the execution times over the real architecture ($T_{exec}$) and the estimated execution times obtained when assigning tasks with the AMTHA algorithm ($T_{est}$) is calculated.

The relative percentage (%$Dif_{rel}$) of difference in $T_{exec}$ is:

$$\% Dif_{rel} = \frac{T_{exec} - T_{est}}{T_{exec}} * 100 \qquad (4)$$

As the volume of communications (or the size of the transmitted packages) between tasks increases, so does the error as a function of the available cache in each core.

In the tests carried out with 8 cores this value was never above 4% and with 64 cores it increases up to 6% (always in applications with a much greater processing load than communication load).

## 7. Conclusions and research lines

Adaptability of MPAHA model and AMTHA algorithm to multicore cluster architectures was analyzed, with good results.

The operation of the AMTHA mapping algorithm was tested over two different multicore cluster architectures showing that it's capable of successfully estimating the execution time of the applications over the multicore architecture (error < 6%) [18].

As regards related future research lines:

- Scalability with long messages (exceeding the capacity of the shared memories).
- Changing the range of parameters for the applications (in particular increasing the number of subtasks)
- Extending tests to 128 cores (adding other 8 blades to the BL260).

An open research line is considering hybrid programming models (integrating message passing and shared memory).

## References

[1] T. W. Burger, "Intel Multi-Core Processors: Quick Reference Guide", http://cachewww.intel.com/cd/00/00/23/19/231912_231912.pdf

[2] M. C. Michael, "Programming models for scalable multicore programming", http://www.hpcwire.com/features/17902939.html, 2007.

[3] L. Chai, Q. Gao, D. K. Panda, "Understanding the Impact of Multi-Core Architecture in Cluster Computing: A Case Study with Intel Dual-Core System", IEEE International Symposium on Cluster Computing and the Grid 2007 (CCGRID 2007), 2007, pp. 471-478.

[4] O. Sinnen, "Task scheduling for parallel systems", Wiley-Interscience, 2007.

[5] A. Grama, A. Gupta, G. Karypis, V. Kumar, "An Introduction to Parallel Computing. Design and Analysis of Algorithms. 2nd Edition", Pearson Addison Wesley, 2003.

[6] A. Kalinov, S. Klimov, "Optimal Mapping of a Parallel Application Processes onto Heterogeneous Platform", 19th

IEEE International Parallel and Distributed Processing Symposium (IPDPS'05), IEEE CS Press, 2005, pp 123.

[7] C. Leopold, "Parallel and Distributed Computing. A survey of Models, Paradigms, and Approaches", Wiley, 2001.

[8] H. Attiya, J. Welch, "Distributed Computing: Fundamentals, Simulations, and Advanced Topics. 2nd Edition", Wiley-IEEE, 2004.

[9] H. Topcuoglu, S. Hariri, M. Wu, "Performance-Effective and Low-Complexity Task Scheduling for Heterogeneous Computing", IEEE Transactions on Parallel and Distributed Systems, Vol. 13, 2002, pp. 260-274.

[10] Goldman, "Scalable Algorithms for Complete Exchange on Multi-Cluster Networks", CCGRID'02, IEEE/ACM, 2002, pp. 286 – 287.

[11] C. Roig, A. Ripoll, M. A. Senar, F. Guirado, E. Luque, "Modelling Message-Passing Programas for Static Mapping", Euromicro Workshop on Parallel and Distributed Processing (PDP'00), IEEE CS Press, 1999, pp. 229-236.

[12] C. Roig, A. Ripoll, M. Senar, F. Guirado, E. Luque, "Exploiting knowledge of temporal behavior in parallel programs for improving distributed mapping". EuroPar 2000, LNCS, Vol. 1900, 2000, pp. 262-71.

[13] L. De Giusti, F. Chichizola, M. Naiouf, A. De Giusti, "Mapping Tasks to Processors in Heterogeneous Multiprocessor Architectures: The MATEHa Algorithm", International Conference of the Chilean Computer Science Society (SCCC 2008), IEEE CS Press, 2008, pp. 85-91.

[14] L. De Giusti, "Mapping sobre Arquitecturas Heterogéneas", Ph.D. thesis, Universidad Nacional de La Plata, La Plata, Argentina, 2008.

[15] J. Cuenca, D. Gimenez, J. Martinez, "Heuristics for Work Distribution of a Homogeneous Parallel Dynamic Programming Scheme on Heterogeneous Systems", 3rd International Workshop on Algorithms, Models and Tools for Parallel Computing on Heterogeneous Networks (HeteroPar'04), IEEE CS Press, 2004, pp. 354-361.

[16] J. C. Cunha, P. Kacsuk, S. Winter, "Parallel Program development for cluster computing: methodology, tools and integrated environments". Nova Science Pub., New York, 2001.

[17] S. Siddha, V. Pallipadi, A. Mallick, "Process Scheduling Challenges in the Era of Multicore Processors", Intel Technology Journal, Vol. 11, No. 4, 2007.

[18] L. De Giusti, F. Chichizola, M. Naiouf, A. De Giusti, E. Luque, "Use of AMTHA in clusters multicores". Technical Report, III-LIDI, 2009.

**Laura De Giusti** received her PhD degree from the Universidad Nacional de La Plata (UNLP) in 2008. She is currently a researcher and assistant professor in the Instituto de Investigación en Informática LIDI (III-LIDI) of the Computer Science School in UNLP. Her main research interests include parallel systems, models and mapping algorithms.

**Franco Chichizola** is a PhD student and researcher assistant in the Instituto de Investigación en Informática LIDI (III-LIDI) of the Computer Science School in UNLP. His main research interests include parallel algorithms and performance analysis.

**Marcelo Naiouf** received his PhD degree from the Universidad Nacional de La Plata (UNLP) in 2004. He is currently a researcher and chair professor in the Instituto de Investigación en Informática LIDI (III-LIDI) of the Computer Science School in UNLP. His main research interests include parallel and distributed systems, algorithms, performance analysis and load planification.

**Armando De Giusti** has university degrees in Electronic Engineering and Computer Science. from the Universidad Nacional de La Plata (UNLP) in 1973. He is currently the head of the Instituto de Investigación en Informática LIDI (III-LIDI) of the Computer Science School in UNLP and CONICET Main Researcher. His research interests include concurrency, distributed and parallel processing, grid computing, real time systems, and computer technology applied to education.

**Emilio Luque** received his PhD degree from the Universidad Complutense de Madrid (UCM) in 1973. He is currently University Full Professor in Universidad Autónoma de Barcelona (UAB), and he is the head of the Computer Architecture and Operating Systems (CAOS) of the UAB. His main research interests include parallel computer architecture, communications, load planification, heterogeneous distributed systems, VoD, and parallel simulation.